\documentclass{tlp}

\usepackage{epsfig}
\usepackage{wrapfig}
\usepackage{subfigure}
\usepackage{rotating}
\usepackage{moreverb}
\usepackage{fancyvrb}

\newcommand{\chr}{{CHR}}

\newcommand{\simp}{{\; \Leftrightarrow \;}}

\newcommand{\prop}{{\; \Rightarrow\; }}
\newcommand{\com}{\rule{0.5pt}{7pt}}

\begin{document}

\title[Union-Find in CHR]{\bf Optimal Union-Find in Constraint Handling Rules}
\author[Tom Schrijvers and Thom Fr{\"u}hwirth]{Tom Schrijvers\thanks{Research Assistant of the fund for Scientific Research - Flanders (Belgium)(F.W.O. - Vlaanderen)}\\
Department of Computer Science, K.U.Leuven, Belgium\\
\email{www.cs.kuleuven.ac.be/\~{}toms/}
\and
Thom Fr{\"u}hwirth\\
Faculty of Computer Science, University of Ulm, Germany\\
\email{www.informatik.uni-ulm.de/pm/mitarbeiter/fruehwirth/}
}


\maketitle

\begin{abstract}
Constraint Handling Rules (CHR) is a committed-choice rule-based language that
was originally intended for writing constraint solvers. In this paper we show
that it is also possible to write the classic union-find algorithm and variants
in CHR.  The programs neither compromise in declarativeness nor efficiency.  We
study 
the time complexity of our programs: 
they match the almost-linear complexity of the best known imperative
implementations. This fact is illustrated with experimental results.

To appear in Theory and Practice of Logic Programming (TPLP).
\end{abstract}


\section{Introduction}

When a new programming language is introduced, sooner or later the question arises
whether classical algorithms can be implemented in an efficient and elegant way.
For example, one often hears the argument that in Prolog some graph algorithms
cannot be implemented with best known complexity because Prolog lacks
destructive assignment that is needed for efficient update of the graph data
structures.
In particular, it is not clear whether the union-find algorithm can be implemented
with optimal complexity in pure (i.e. side-effect-free) Prolog \cite{McAllester}.

In this programming pearl, we give a positive answer for the Constraint Handling
Rule (CHR) programming language. We give a CHR implementation with the best known
worst case and amortized time complexity for the classical union-find algorithm with
path compression for find and union-by-rank. This is particularly remarkable,
since originally CHR was intended for implementing constraint solvers only.

CHR is a concurrent committed-choice constraint logic programming
language consisting of guarded rules that transform multi-sets of constraints
(atomic formulae).
In \chr, one distinguishes two main kinds of rules: Simplification rules replace
constraints by simpler constraints, e.g.\
{\tt X\(\geq\)Y \(\land\) Y\(\geq\)X \(\simp\) X\(=\)Y}. Propagation rules add new
constraints, which may cause further
simplification, e.g.\ {\tt X\(\geq\)Y\(\land\)Y\(\geq\)Z\(\prop\)X\(\geq\)Z}.
This combination of propagation and multi-set transformation of logical formulae
make CHR a unique and powerful declarative programming language.

Closest to our work is the presentation of a logical algorithm for the
union-find problem in \cite{McAllester}. In a hypothetical bottom-up inference
rule language with permanent deletions and rule priorities, a set of
rules for union-find is given that is proven to run in $\mathcal{O}(M + N
log(N))$ worst-case time for a sequence of $M$ operations on $N$ elements. The
direct efficient implementation of these inference rule system seems 
infeasible. 
It is also not clear whether the rules given in \cite{McAllester} describe the
standard union-find algorithm as can be found in text books such as
\cite{Cormen}. The authors remark that giving a rule set with optimal amortized
complexity is complicated. 

In contrast, we give an executable and efficient
implementation that directly follows the pseudo-code presentations found in text
books and that also has optimal amortized complexity. Moreover, we do not 
rely on rule priorities. 

This paper is structured as follows. In the next Section, we review the
classical union-find algorithms. Constraint Handling Rules (CHR) are briefly
presented in Section \ref{sec:chr}.  Then, in Section \ref{sec:ufd_basic} we
present the implementation of the classical union-find algorithm in
CHR.
 An improved version
of the implementation, featuring path compression and union-by-rank, is
presented next in Section \ref{sec:ufd_rank}. In Section \ref{sec:complexity},
we argue that this implementation has the same time complexity as the best known
imperative implementations. This claim is experimentally evaluated in Section
\ref{sec:evaluation}. Finally, Section \ref{sec:conclusion} concludes.


\section{The Union-Find Algorithm}\label{sec:unifionfind}

The classical union-find (also: disjoint set union) algorithm was introduced
by Tarjan in the seventies \cite{Tarjan}. 
A classic survey on the topic is \cite{Galil}.
The algorithm solves the problem of maintaining a collection of disjoint sets
under the operation of union.  Each set is represented by a rooted tree, whose
nodes are the elements of the set. The root is called the {\em representative}
of the set. The representative may change when the tree is updated by a union
operation. With the algorithm come three operations on the sets:
\begin{itemize}
\item \texttt{make(X)}: create a new set with the single element \texttt{X}.
\item \texttt{find(X)}: return the representative of the set in which \texttt{X}
  is contained.
\item \texttt{union(X,Y)}: join the two sets that contain \texttt{X} and
  \texttt{Y}, respectively (possibly destroying the 
old sets and changing the representative).
\end{itemize}
A new element must be introduced exactly once with \texttt{make} before being
subject to \texttt{union} and \texttt{find} operations.

In the naive algorithm, these three operations are implemented as follows.
\begin{itemize}
\item \texttt{make(X)}: generate a new tree with the only node \texttt{X},
  i.e. \texttt{X} is the root.
\item \texttt{find(X)}: follow the path from the node \texttt{X} to the root of
  the tree. Return the root as representative.
\item \texttt{union(X,Y)}: find the representatives of \texttt{X} and \texttt{Y},
  respectively. To join the two trees, it suffices to \texttt{link} them by
  making one root point to the other root.
\end{itemize}

This following imperative pseudo-code implements this algorithm:
\medskip
\begin{Verbatim}[frame=single,framesep=5mm,commandchars=\\\{\},codes={\catcode`$=3},label=Naive Union-Find Algorithm]
make(x) 
  p[x] $\leftarrow$ x

union(x,y)
  link(find(x),find(y))

link(x,y) 
  {\bf if} x $\neq$ y
        {\bf then} p[y] $\leftarrow$ x

find(x) 
  {\bf if} x $\neq$ p[x]
        {\bf then} {\bf return} find(p[x])
        {\bf else} {\bf return} x
\end{Verbatim}

In this pseudo-code \texttt{p[x]} denotes the ancestor of \texttt{x} in the tree.
If \texttt{x} is the root, then \texttt{p[x]} equals \texttt{x}.

The naive algorithm requires $\mathcal{O}(N)$ time per \texttt{find} (and \texttt{union}) in the
worst case, where $N$ is the number of elements (\texttt{make} operations).  With two
independent optimizations that keep the tree shallow and balanced, one can
achieve quasi-constant (i.e. almost constant) {\em amortized} running time per
operation.

The first optimization is {\em path compression} for find. It moves nodes closer
to the root. After \texttt{find(X)} returned the root of the tree,
we make every node on the path from \texttt{X} to the root point directly to the
root.
The second optimization is {\em union-by-rank}. It keeps the tree shallow by
pointing the root of the smaller tree to the root of the larger tree.
{\em Rank} refers to an upper bound of the tree depth.
If the two trees have the same rank, either direction of pointing is chosen but
the rank is increased by one. 

For each optimization alone and for using both of them together, the worst case
time complexity for a single find or union operation is $\mathcal{O}(log(N))$. For a
sequence of $M$ operations on $N$ elements, the worst complexity is $\mathcal{O}(M + N
log(N))$. When both optimizations are used, the amortized complexity is
quasi-linear, $\mathcal{O}(M + N \alpha(N))$, where $\alpha(N)$ is an inverse of the
Ackermann function and is less than 5 for all practical $N$.

In the naive pseudo-code, the \texttt{make}, \texttt{link} and \texttt{find}
operations have to be redefined as follows, to add union-by-rank and path
compression.
\medskip
\begin{Verbatim}[frame=single,framesep=5mm,commandchars=\\\{\},codes={\catcode`$=3},label=Union-Find with Union-by-Rank and Path Compression]
make(x) 
  p[x] $\leftarrow$ x
  rank[x] $\leftarrow$ 0

link(x,y)
  {\bf if} x $\neq$ y
    {\bf if} rank[x] $\geq$ rank[y]
      {\bf then} p[y] $\leftarrow$ x
           rank[x] $\leftarrow$ max(rank[x],rank[y] + 1)   
      {\bf else} p[x] $\leftarrow$ y

find(x)
  {\bf if} x $\neq$ p[x]
        {\bf then} p[x] $\leftarrow$ find(p[x])
  {\bf return} p[x]
\end{Verbatim}

The union-find algorithm has applications in graph theory (e.g. efficient
computation of spanning trees).  
By definition of set operations, a union operator working on representatives
of sets is an equivalence relation, i.e. we can view sets as equivalence
classes.
When the union-find algorithm is extended to deal with nested terms to perform
congruence closure, the algorithm can be used for term unification in theorem
provers and in Prolog.\footnote{It is straightforward to combine the existing
CHR solvers for term unification with our union-find implementation.}
The WAM \cite{wam}, Prolog's traditional abstract machine, uses the basic 
version of union-find for variable aliasing.
While {\em variable shunting}, a limited form of path compression, is used in
some Prolog implementations \cite{VariableShunting}, we do not know of
any implementation of the optimized union-find that keeps track of
ranks or other weights.

\section{Constraint Handling Rules (CHR)}\label{sec:chr}

In this section we give an overview of the syntax and operational semantics for
constraint handling rules (CHR)~\cite{Fru98,book,refined}.

\subsection{Syntax of \chr}

We use two disjoint sets of predicate symbols for two different
kinds of constraints: {built-in (pre-defined) constraint symbols} 
which are solved by a given constraint solver,
and
{CHR (user-defined) constraint symbols} 
which are defined by the rules in a \chr\ program.
There are three kinds of rules:

\begin{center}
\begin{tabular}{ll}
{\em Simplification rule:} & {\it Name} $\; @ \; H \simp C \; \com \; B,$ \\
{\em Propagation rule:} & {\it Name} $\; @ \; H \prop C \; \com \; B,$ \\
{\em Simpagation rule:} & {\it Name} $\; @ \; H \setminus H' \simp C \; \com \; B,$ \\
\end{tabular}
\end{center}

\noindent where {\em Name} is an optional, unique identifier of a rule, the {\em head}
$H$, $H'$ is a non-empty comma-separated conjunction of CHR constraints, the
{\em guard} $C$ is a conjunction of built-in constraints, and the {\em body} $B$
is a goal. A {\em goal (query)} is a conjunction of built-in and CHR
constraints.
A trivial guard expression ``\texttt{true |}'' can be omitted from a rule.
Simpagation rules abbreviate simplification rules of the form
$\mathit{Name} \; @ \; H, H' \simp C \; \com \; H, B$.

\subsection{Operational Semantics of \chr}\label{sec:chr:semantics}

Given a query, the rules of the program are applied to exhaustion. 
A rule is {\em applicable}, if 
its head constraints are matched by constraints in the current goal
one-by-one 
and if, under this
matching, the guard of the rule is implied by the built-in constraints in the
goal.
Any of the applicable rules can
be applied, and the application cannot be undone, it is committed-choice (in
contrast to Prolog).
When a simplification rule is applied, the matched constraints in the current
goal are replaced by the body of the rule, when a propagation rule is applied,
the body of the rule is added to the goal without removing any constraints.

This high-level description of the operational semantics of CHR leaves two main
sources of non-determinism: the order in which constraints of a query are
processed and the order in which rules are applied. As in Prolog, almost all CHR
implementations execute queries from left to right and apply rules top-down in
the textual order of the program.\footnote{The nondeterminism due to 
the wake-up order of delayed constraints and multiple matches for the same rule
are of no relevance for the programs discussed here.} This behavior has
been formalized in the so-called refined semantics that was also proven to be a
concretization of the standard operational semantics \cite{refined}.

In this refined semantics of actual implementations, a CHR constraint in a query
can be understood as a procedure that goes efficiently through the rules of the program
in the order they are written,
and when it matches a head constraint of a rule, it will look for the other,
{\em partner
constraints} of the head in the {\em constraint store} and check the guard until an
applicable rule is found. We consider such a constraint to be {\em active}. If the
active constraint has not been removed after trying all rules, it will be put
into the constraint store. Constraints from the store will be reconsidered
(woken) if newly added built-in constraints constrain variables of the
constraint, because then rules may become applicable since their guards
are now implied. Obviously, ground constraints need never to be considered for
waking.
The discussion above will be of use in Section \ref{sec:complexity} where we
derive the time complexities of our CHR programs.

\section{Implementing Union-Find in CHR}\label{sec:ufd_basic}

The following CHR program in concrete ASCII syntax implements the operations and
data structures of the naive union-find algorithm without optimizations.
\begin{itemize}
\item {\tt root(X)}   represents {\tt p[X] = X},
\item {\tt X \~{}> PX}   represents {\tt p[X] = PX},
\item {\tt find(X,R)} implements {\tt R = find(X)},
\item {\tt make} and {\tt union} are identical.
\end{itemize}
The constraints \texttt{make/1}, \texttt{union/2}, \texttt{find/2} and
  \texttt{link/2} define the operations, so we call them {\em
operation constraints}. 
The constraints \texttt{root/1} and \texttt{\~{}>/2} represent the tree data
structure and we call them {\em data constraints}.
\medskip
\begin{Verbatim}[frame=single,framesep=5mm,commandchars=+\{\},codes={\catcode`$=3},label=ufd\_basic]
make     @ make(X) <=> root(X). 

union    @ union(X,Y) <=> find(X,A), find(Y,B), link(A,B). 

findNode @ X ~> PX \ find(X,R) <=> find(PX,R). 
findRoot @ root(X) \ find(X,R) <=> R=X. 

linkEq   @ link(X,X) <=> true. 
link     @ link(X,Y), root(X), root(Y) <=> Y ~> X, root(X). 
\end{Verbatim}

The elements we apply \texttt{union} to are constants as usual for union-find
algorithms.  Hence the arguments of all constraints are constants, with the
exception of the second argument of \texttt{find/2} that must be a variable
that is bound to a constant in the rule \texttt{findRoot}.

Actually, the use of the built-in constraint \texttt{=} in this rule is
restricted to returning the element \texttt{X} in the parameter \texttt{R}. In
particular no full unification is ever performed. 

This union-find program and the one in the next section are not\linebreak
confluent~\cite{CW389,wclp2005}, i.e. results are dependent of the order in which
applicable rules are applied. The main reason is that the relative order of {\tt
find} and {\tt union} operations matters for the outcome of a {\tt find}. This
behavior is inherent in the union-find algorithm due to its update of the tree
structure (see also the discussion of the logical reading of the rules
in~\cite{CW389}).

\section{Optimized Union-Find}\label{sec:ufd_rank}

The following CHR program implements the optimized classical union-find Algorithm with
path compression for find and union-by-rank \cite{Tarjan}. The \texttt{union/2} constraint
is implemented exactly as for the naive algorithm.

\medskip
\begin{Verbatim}[frame=single,framesep=1mm,commandchars=*\{\},codes={\catcode`$=3},label=ufd\_rank]
make      @ make(X) <=> root(X,0). 

union     @ union(X,Y) <=> find(X,A), find(Y,B), link(A,B). 

findNode  @ X ~> PX , find(X,R) <=> find(PX,R), X ~> R. 
findRoot  @ root(X,_) \ find(X,R) <=> R=X. 

linkEq    @ link(X,X) <=> true. 
linkLeft  @ link(X,Y), root(X,RX) root(Y,RY) <=> RX >= RY | 
               Y ~> X, NRX is max(RX,RY+1), root(X,NRX). 
linkRight @ link(X,Y), root(Y,RY), root(X,RX) <=> RY >= RX | 
               X ~> Y, NRY is max(RY,RX+1), root(Y,NRY). 
\end{Verbatim}

When compared to the naive version \texttt{ufd\_basic}, we see that
\texttt{root} has been extended with a second argument that holds the rank of
the root node.
The rule \texttt{findNode} has been extended for path compression already
during the first pass along the path to the root of the tree. This is achieved
by the help of the variable \texttt{R} that serves as a place holder for
the result of the find operation.
The \texttt{link} rule has been split into two rules, \texttt{linkLeft} and
\texttt{linkRight}, to reflect the optimization of union-by-rank: The smaller
ranked tree is added to the larger ranked tree without changing its rank. When
the ranks are the same, either tree is updated (both rules are applicable) and
the rank is incremented by one.

\section{Time Complexity}\label{sec:complexity}

While automatic complexity analysis results for CHR exist
\cite{chr_complexity}, these do not take into account the refined operational
semantics of CHR \cite{refined} and hence will yield only a very crude
approximation of the actual complexity.

Instead we establish the time complexity of our CHR programs by first
showing that they are operationally equivalent to the respective imperative
algorithms.  By showing next that all the individual computation steps in
the CHR program have the same complexity as their imperative counterparts,
we have then effectively proven that the overall time complexity properties
are identical to the ones of imperative implementations.

\subsection{Operational Equivalence}

We start with considering the naive algorithm.
Because of the refined operational semantics of CHR, the query of
\texttt{make/1}, \texttt{union/2} and \texttt{find/2} constraints (and any
other conjunction of constraints) is evaluated from left to right, just as
is the case for equivalent calls for the imperative program.

Because of this execution order, the operation constraints behave just as
their imperative counterparts. The
imperative \texttt{if-then} and \texttt{if-then-else} constructs are encoded
as multiple rules. The appropriate rule will be chosen because of a combination
of different matchings, partner constraints and guards.

Moreover, the recursion depth for
the \texttt{find/2} constraint is equal to the path from the initial node
to the root just like in the imperative algorithm.
The unification in the body of the {\tt findRoot} rule cannot wake up any
constraints, since the variable that is bound to a constant does not occur
in any other constraint processed so far.

It is clear from the CHR program and the refined operational semantics, that
there is only ever at most one operation constraint in the constraint store.
Moreover, whenever a data constraint is called, the operation constraint
has already been removed. Thus a data constraint will never trigger any rule,
because of lack of the necessary partner constraint.

\subsection{Time Complexity Equivalence}

Now that we have shown the operational equivalence of the CHR program with
the imperative algorithm, we still need to show that the time complexities of
the different computation steps (corresponding to rule applications) are also
equal.

The following time complexity assumptions of a CHR implementation are reasonable
(based on the discussion of CHR operational semantics in Section
\ref{sec:chr:semantics}). They are effectively implemented by the SICStus
\cite{sicstus}, HAL \cite{duck:optimizing} and K.U.Leuven \cite{kulchr} CHR systems. All
of the following operations of the refined operational semantics take constant
time:
\begin{itemize}
\item The \texttt{Activate} transition, excluding the cost of adding the constraint to the constraint store. 
\item The \texttt{Drop} transition, i.e. ending the execution of a constraint.
\item The \texttt{Default} transition, i.e. switching from trying one rule to trying the next rule.
\item Matching for Herbrand variables and constants,
      given a bounded reference chain length. This occurs in the \texttt{Simplify} and
      \texttt{Propagate} transitions.
\item Instantiating a variable that does not occur in any constraints, i.e.
      an obvious optimization of the \texttt{Solve} transition.
\item Checking simple arithmetic built-in constraints like \texttt{>=} and {\tt min}.
\end{itemize}


The following complexity assumptions can be realized in practice by
appropriate indexing, i.e. constraint store lookup based on shared variables.
\begin{enumerate}
\item The cost of finding all constraints with a particular value
      in a particular argument position is constant. 
      Even if there are no such constraints.
      The cost of obtaining one by one all constraints from such a set is 
      is proportional to the size of the set. 
\item The CHR constraint store allows constant addition and 
      deletion of any constraint.
\item If more than one partner constraint has to be found, 
      an ordering of look-ups is preferred, if possible, such that the next
      constraint to look-up shares a variable with the previously
      found constraints and the active constraint. 
\end{enumerate}
The last item is a heuristic presented in
\cite{duck:optimizing} and implemented in the HAL and KULeuven CHR systems. In
Section \ref{sec:evaluation} we will discuss appropriate constraint store
data structures that fulfill the remaining assumptions.

From these assumptions it is clear that processing a data constraint takes
constant time: the constraint is called, some rules are tried,
some partner constraints which share a variable with the active constraint
are looked for, but none are present, and finally the call ends
with inserting the data constraint into the constraint store.

Hence our naive CHR implementation has the same time complexity properties
as the naive imperative algorithm.

The proof of operational equivalence and equivalent complexity of the
optimized algorithm and CHR program is similar.
Because of this equivalence with the imperative algorithm, our CHR program
also has worst-case time complexity is $\mathcal{O}(M + N log(N))$ and
amortized time complexity $\mathcal{O}(M + N \alpha(N))$.

\section{Experimental Evaluation}\label{sec:evaluation}

To experimentally validate the derived complexity derived above, we have run the
CHR program in SWI-Prolog \cite{SWI} using the K.U.Leuven CHR system \cite{kulchr}. 
This CHR system will use hashtables as constraint stores for lookups on shared
variables that are ground. These hashtables allow for efficient lookup,
insertion and deletion of constraints.  By adding the appropriate mode
declarations to our program, the system establishes the groundness of shared
variables.

By initializing the hashtables to the appropriate sizes and choosing
the used constants appropriately, it is possible to avoid hashtable
collisions. Then, the hashtables essentially behave as arrays (just as in
the imperative code) and the assumptions of the previous section are
effectively realized.

In contrast, the first and de facto standard CHR system, available in
SICStus \cite{sicstus}, does not provide the necessary constant time
operations. While it does have constant lookup time for all constraint
instances of a particular constraint that contain a particular variable,
it does not distinguish between argument positions. Hence, the lookup of
\texttt{root(X,R)} can be done in constant time given \texttt{X}, but the
lookup of \texttt{X \~{}> Y} is proportional to the number of \texttt{\~{}>}
constraints \texttt{X} appears in. If \texttt{X} is a node with $K$
children, then it will be $\mathcal{O}(K)$.  Moreover, while the insertion
of a constraint instance is $\mathcal{O}(1)$, deletion is $\mathcal{O}(I)$,
where $I$ is the total number of instances of the constraint.  

\begin{figure}[h!]
\begin{center}
        \subfigure[SICStus and SWI-Prolog Array\label{figure:sicstus}]{\includegraphics[width=0.9\textwidth]{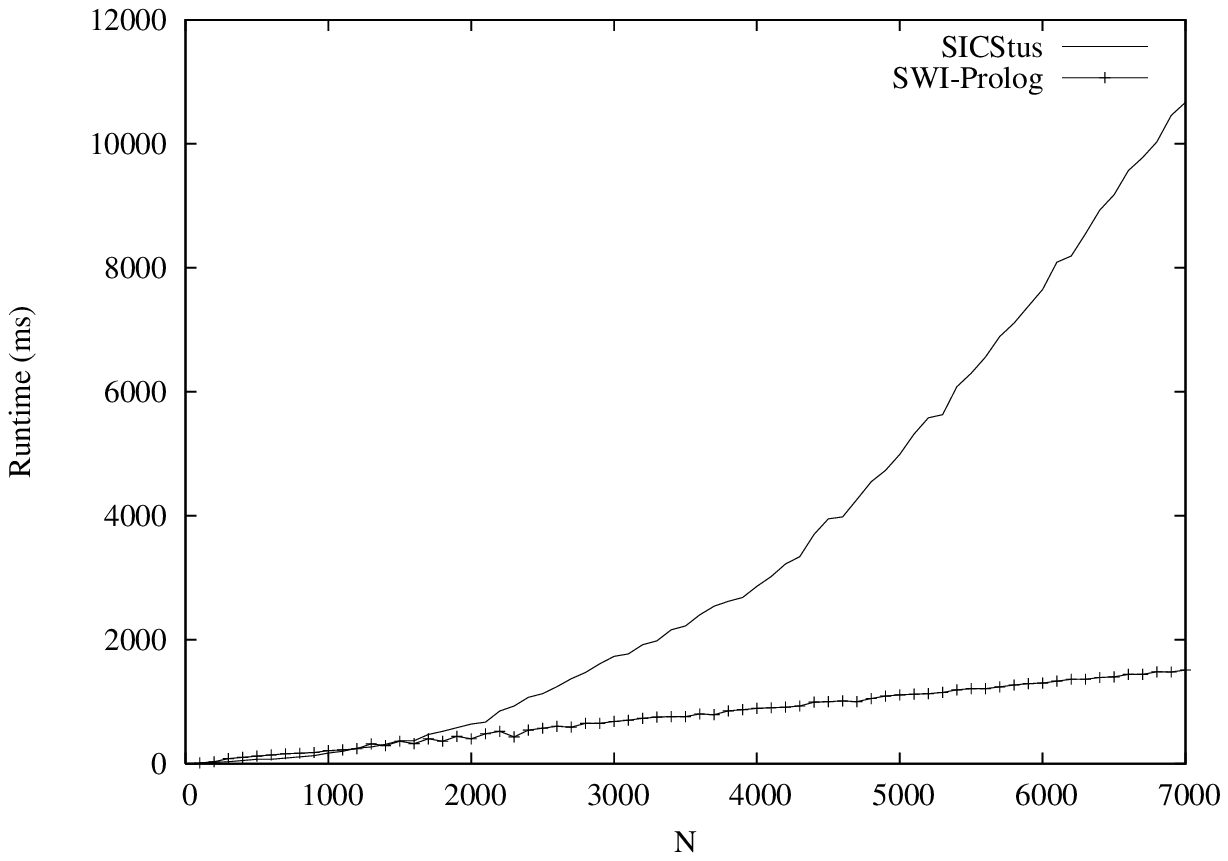}}
        \subfigure[Detail of Figure \ref{figure:sicstus}: SWI-Prolog Array\label{figure:swi:array}]{\includegraphics[width=0.5\textwidth]{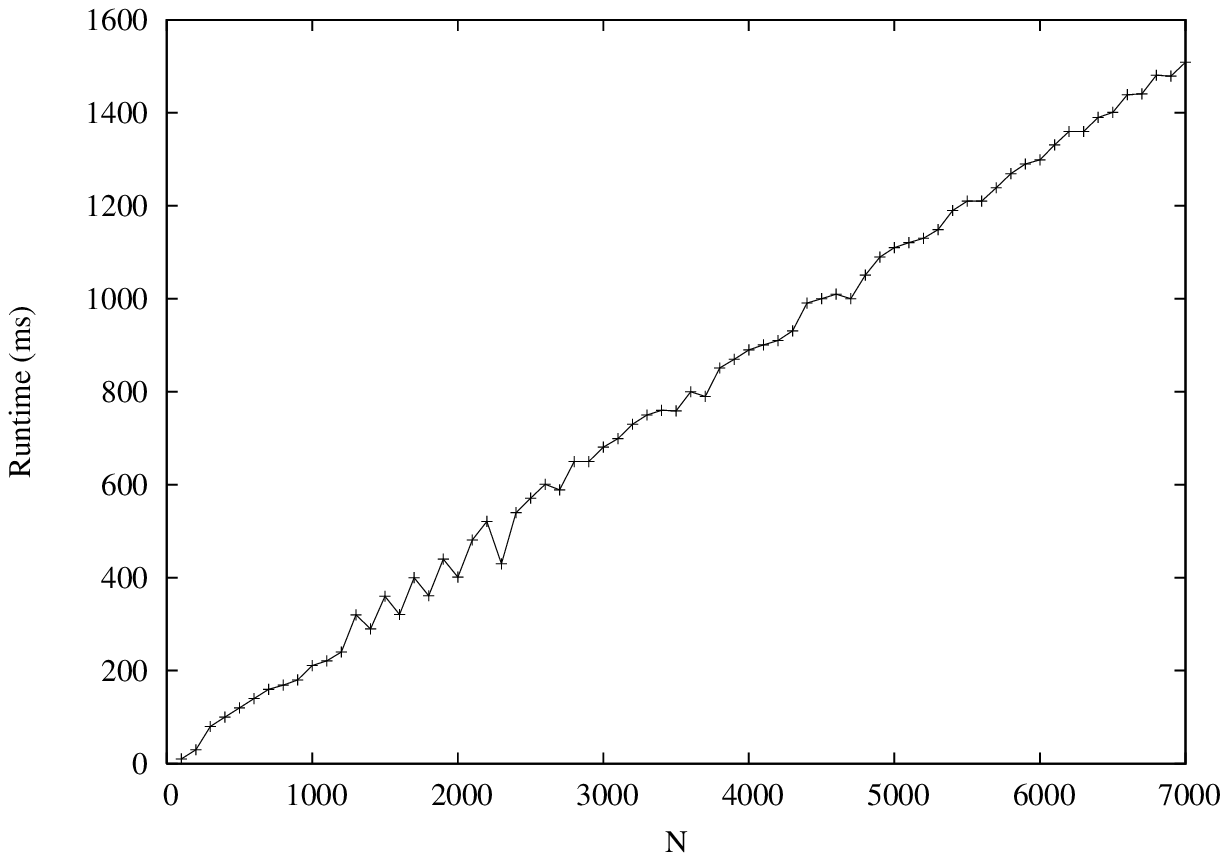}}
        \subfigure[SWI-Prolog Hashtable\label{figure:swi:hashtable}]{\includegraphics[width=0.5\textwidth]{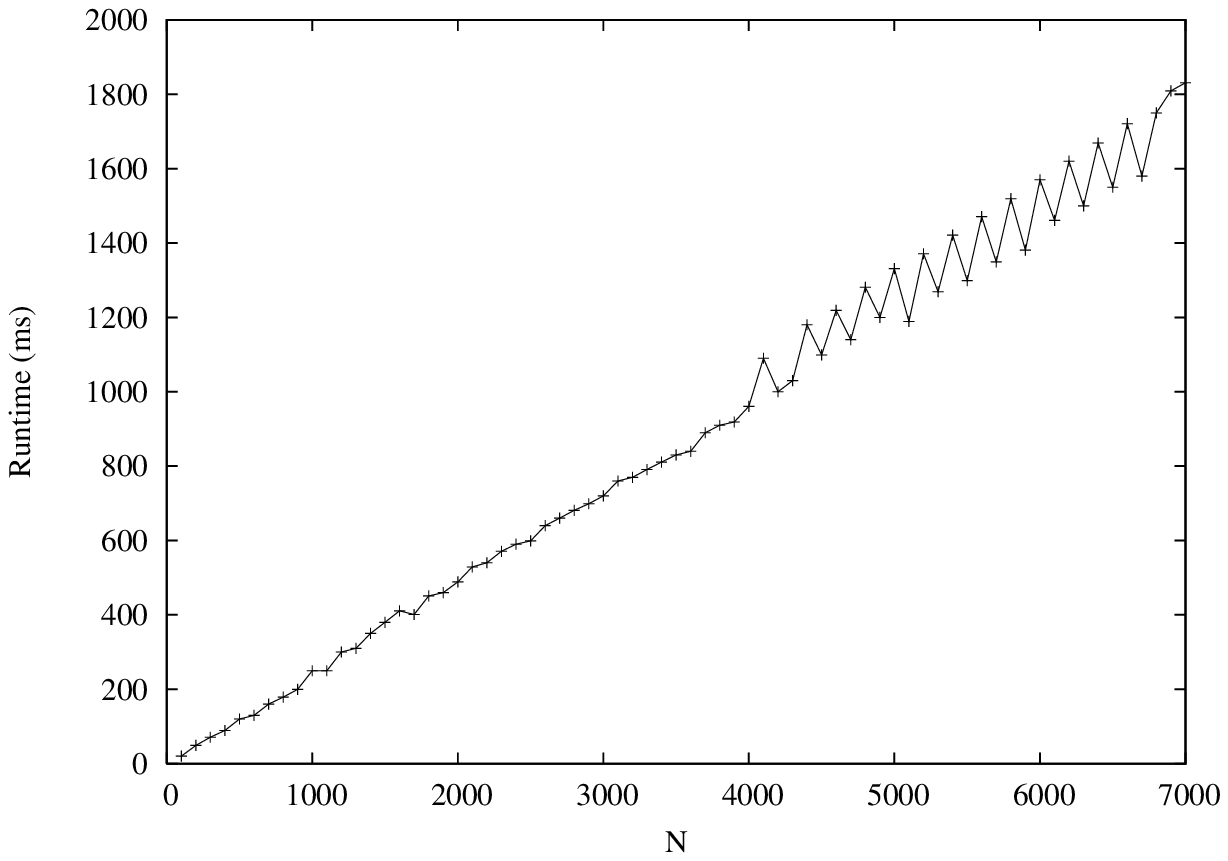}}
\end{center}
\caption{Observation of behavior for contrived unions.}\label{fig:complexity:contrived}
\end{figure}

The queries we use in our experimental evaluation consist of $N$ calls to
\texttt{make/1}, to create $N$ different elements, followed by $N$ calls to
\texttt{union/2} and $N$ calls to \texttt{find/2}. The input arguments of the
latter two are chosen at random among the elements.  
Even the SICStus CHR system exhibits near-linear behavior for a
random set of \texttt{union} operations. So we consider instead a
contrived set of \texttt{union} operations: disjoint trees of elements
are unioned pairwise until all elements are part of the same tree. Figures
\ref{figure:sicstus} and \ref{figure:swi:array} show the runtime results for SICStus and
SWI-Prolog. It is clear from the figure that SICStus does not show the
optimal quasi-linear behavior anymore which is still observed in SWI-Prolog.

We also compare the above two cases to the case where the hashtables are
not initialized to a large enough size, but instead double in size and
rehash each time their load equals their size. While individual hashtable
operations no longer take constant time, on average they do \cite{Cormen},
which is sufficient for our complexity analysis. This is confirmed by
experimental evaluation (see Figure \ref{figure:swi:hashtable}).

The above comparisons illustrate that it is vital for efficiency to use a
CHR system with the proper constraint store data structures. To the best of
our knowledge, the K.U.Leuven CHR system is currently the only system that
provides hashtable-based indexing constraint stores.


\section{Conclusion}\label{sec:conclusion}

We have shown in this paper that it is possible to implement the classical
union-find algorithm concisely and efficiently in constraint handling rules
(CHR). The implementation is easily extended with optimizations like
path-compression and union-by-rank. In addition, we showed the optimal 
time complexity properties of our implementations. 
The declarative nature of CHR is no compromise for time complexity.

At \texttt{http://www.cs.kuleuven.ac.be/\~{}toms/Research/CHR/UnionFind/}
all\linebreak presented programs as well as related material are 
available for download.  The programs can be run with the proper time
complexity in the latest release of SWI-Prolog.
The technical report \cite{CW389} associated with this paper contains a
detailed analysis of the confluence properties and logical semantics of our
union-find implementations.

In future work we intend to investigate implementations for other variants of
the union-find algorithm. For a parallel version of the union-find algorithm
parallel operational semantics of CHR have to be investigated (confluence may be
helpful here).  A dynamic version of the algorithm, e.g. where unions can be
undone, would presumably benefit from dynamic CHR constraints as defined in
\cite{ArminWolf2}.

{\bf Acknowledgements.} We would like to thank the participants of the first
workshop on CHR in May 2004 for raising our interest in the subject. Marc Meister and the
students of the constraint programming course at the University of Ulm in summer
2004 helped by implementing and discussing their versions of the union-find
algorithm.
Part of this work was done while Tom Schrijvers was visiting the
University of Ulm in November 2004.
Last but not least we would like to thank our reviewers for their suggestions
and comments that greatly helped to improve the paper.


\bibliographystyle{acmtrans} 
\bibliography{paper}


\end{document}